# Unconventional electro-mechanical response in Ferrocene assisted gold atomic chain

Biswajit Pabi[a,$,*], Štěpán Marek[b,c], Tal Klein[d], Arunabha Thakur[e], Richard Korytár[b] and Atindra Nath Pal[a]*

[a]Department of Condensed Matter and Materials Physics, S. N. Bose National Centre for Basic Sciences, Sector III, Block JD, Salt Lake, Kolkata 700106, India.
[b]Department of Condensed Matter Physics, Faculty of Mathematics and Physics, Charles University, CZ-121 16 Praha 2, Czech Republic.
[c]Institute of Theoretical Physics, University of Regensburg, Universitätsstraße 31, D-93053, Regensburg, Germany.
[d]Department of Chemical and Biological Physics, Weizmann Institute of Science, Rehovot 7610001, Israel.
[e]Department of Chemistry, Jadavpur University, Kolkata-700032, India

[$]Present address: Institut für Experimentelle und Angewandte Physik, Christian-Albrechts-Universität, 24098 Kiel, Germany.

*Email: biswajitpabi1@gmail.com, atin@bose.res.in

**Abstract:**

Atomically thin metallic chains serve as pivotal systems for studying quantum transport, with their conductance strongly linked to the orbital picture. Here, we report a non-monotonic electro-mechanical response in a gold-ferrocene junction, characterized by an unexpected conductance increase over a factor of ten upon stretching. This response is detected in the formation of ferrocene-assisted atomic gold chain in a mechanically controllable break junction at a cryogenic temperature. DFT based calculations show that tilting of molecules inside the chain modifies the orbital overlap and the transmission spectra, leading to such non-monotonic conductance evolution with stretching. This behavior, unlike typical flat conductance plateaus observed in metal atomic chains, pinpoints the unique role of conformational rearrangements during chain elongation. Our findings provide a deeper understanding of the role of orbital hybridization in transport properties and offer new opportunities for designing nanoscale devices with tailored electro-mechanical characteristics.

**Introduction:**

Break junction experiments highlight the interplay between electrical and mechanical properties in nanoscale junctions through observed conductance behavior during stretching, offering unique control beyond traditional electronics[1–3]. The conductance evolution in these junctions is intricately linked to changes in orbital hybridization driven by conformational rearrangements, manifesting as plateaus with either constant or decaying conductance. While mechanical stretching typically weakens molecule–electrode coupling and increases tunneling distance, leading to reduced conductance, an unexpected increase in conductance has been observed in aluminum atomic junctions, as well as alkane and benzene dithiol molecular junctions. This counterintuitive behavior is attributed to an enhanced local density of states at the Fermi level or force-induced energy level alignment[4–6]. Simulations inspired by these observations further suggest microscopic mechanisms involving the formation of metal atomic chains and changes in metal–molecule bond angles[7,8]. Additionally, anomalous U-shaped conductance plateaus have been predicted for helicene molecules connected by chain of carbon atoms, attributed to non-monotonic variations in the HOMO–LUMO gap during stretching[9].

Observations presented in this letter reveal both increasing and U-shaped plateaus in Au/ferrocene/Au junctions, showcasing conductance behavior that has not been previously observed. Notably, the increasing conductance traces are significant, with the conductance rising by up to an order of magnitude. Clue to this unconventional response lies in the formation of ferrocene assisted gold atomic chains. It is important to note that pure or molecule-assisted chain formation has been observed in metals like Au, Pt, and Ir in their pure forms, as well as in other metals (e.g., Ag, Cu, Fe, Ni, Pd, Co) in the presence of gaseous molecules such as $O_2$, $H_2$, $D_2$, $N_2$, CO, $H_2O$ and even, Pt with more complex molecules like benzene, naphthalene, or anthracene[10–27]. In all the cases, the conductance plateaus were flat or decaying, making our findings a notable exception and introducing organometallic as a new molecule for driving metal atomic chain formation. The correlation between unconventional response and chain formation is elucidated by the first principles calculations, revealing the modifications of metal-molecule coupling via changing of the metal-molecule binding site during molecule assisted chain formation. Structural adaptability, supported by the orbital characteristics of the ferrocene molecule, emerges as the key factor behind this unique electromechanical behavior. Overall, our findings reveal a novel type of electromechanical response in molecular junctions and highlight the critical role of microstructural rearrangements during stretching, with implications for mechano-responsive molecular electronic devices.

**Results and discussions:**

Our study addressing the electro-mechanical response of molecular junction, is based on the ferrocene molecule suspended between two gold (Au) tips. A mechanically controllable break junction (MCBJ) set up[28,29] (**Figure 1**) is used to form in-situ molecular junction by spreading molecules from a locally heated molecular source towards the continuously breaking and making atomic junction at 4.2 K. Direct current (dc) conductance (current\voltage) measurement of the junction is performed at 4.2 K and 77 K as a function of inter-electrode separation (displacement) when the junction is broken and reformed repeatedly, capturing a broad range of configurations. Experimental findings are complemented by theoretical calculations, with methodological details provided in **Section 1** of the **Supplementary Information**.

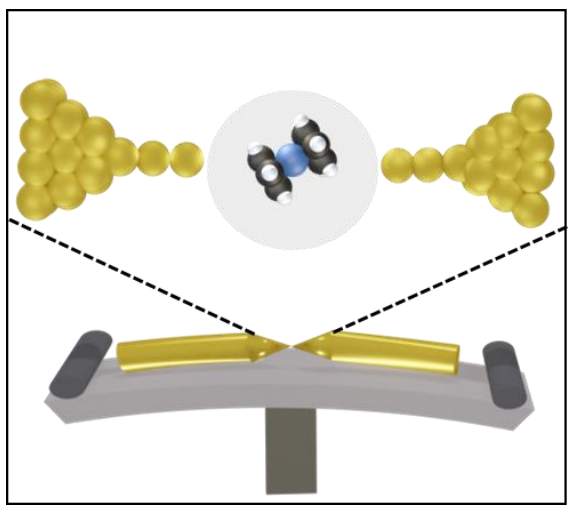

**Figure 1: Experimental set up.** Schematic illustration of the mechanically controllable break junction (MCBJ) set up to form a single molecular junction together with a visual presentation of a molecule linked by chain of metal atoms.

Experiment is started by characterizing the Au junction prior to the insertion of molecules at 4.2 K. **Figure 2a** shows several conductance traces, recorded during the stretching of an Au wire and for which, the conductance decreases in stepwise manner as the contact diameter reduces. Flat conductance plateau at ~1.0 $G_0$ ($G_0=2e^2/h$, the quantum of conductance, where *e* is the electron charge and *h* is the Planck's constant) is attributed to the conductance through a single-atom cross section[30], whereas the exponentially decaying features below 1 $G_0$ are characteristic of tunneling transport. Subsequently, the molecule is sublimated onto the cold junction, which significantly alters the conductance traces, as demonstrated in **Figure 2b**. This confirms the formation of Au/ferrocene/Au junctions. Plateaus with unusual trajectories, exhibiting conductance values from ~ 0.3 $G_0$ to 0.0005 $G_0$ are observed in contrast to conventional flat or decaying plateaus. Furthermore, subsequent plateaus can develop in a single stretching event via jump in conductance. For sake of presentation, we subdivide the conductance values into two regions: < 0.05 $G_0$ as low conductance (LG) and > 0.05 $G_0$ as high conductance (HG). Plateaus can occur in both regions simultaneously (Type-2) or independently (Type-1 for HG and Type-3 for LG); however, their shapes differ between HG and LG. For HG, conductance increases immediately after junction formation, displaying several rising plateaus with sharp jumps in between, forming a tilted 'Z' shape (**Figure 2b**). These 'Z'-shaped features reflect the abrupt changes in the junction's geometry during stretching. In contrast, for LG, conductance initially decreases before increasing, resulting in 'U'-shaped features (**Figure 2b**). These 'U'-shaped features indicate a gradual stretching of the molecule-electrode contact as conductance changes continuously. In fact, this contrasting response between HG and LG plateaus persists even in those cases where both types of plateaus appear simultaneously, as seen in Type-2. It is important to note that the observed increase in conductance as well as the 'U'-shaped features represent an unusual electro-mechanical response. Typically, increasing the distance between two electrodes weakens the metal-molecule interface and increases the tunneling distance, leading to a decrease in conductance. However, an exception to this is observed here. Conductance traces in **Figure 2b** are recorded at a bias voltage of 300 mV. Further experiments at different bias voltages (20 mV, 50 mV, 100 mV, and 200 mV) confirm that this unusual electro-mechanical response is consistent across all voltages, ruling out the influence of electric field. However, the potential impact of electric field on the formation of HG and LG plateaus is observed, which is postponed to the next section.

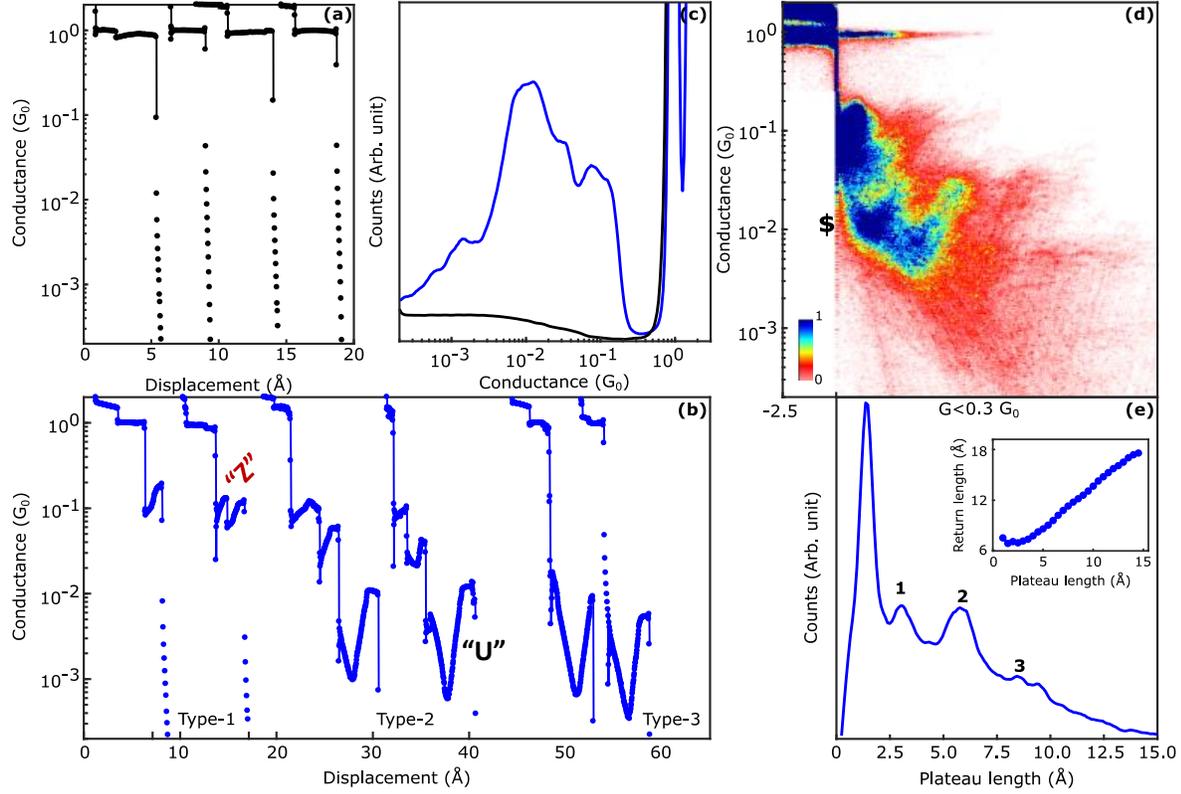

**Figure 2: Conductance characteristics of Au/ferrcene/Au junction at 4.2 K. (a, b)** Examples of conductance-displacement breaking traces of gold (Au) atomic junction before (a) and after (b) insertion of ferrocene molecule. Traces are shifted for clarity. They are categorized into three types based on the presence of plateaus with conductance ~ 0.1 $G_0$ (HG) and 0.01 $G_0$ (LG): Type-1 (only HG), Type-2 (both HG and LG), and Type-3 (only LG). **(c)** 1D conductance histogram of Au (black) and Au/ferrocene/Au junction (blue), prepared from 10000 and 7099 traces using 50 bins per decade. **(d)** 2D conductance-displacement density plot of the same set of traces, generated using 100 bins per decade. Zero displacement is assigned at 0.3 $G_0$, top edge of the peak in the conductance histogram. **(e)** Plateau length histogram of Au/ferrocene/Au junction considering same traces, showing distribution of plateau lengths from zero displacement (0.3 $G_0$) to rupture (0.0005 $G_0$). 200 bins are used to calculate the histogram. Inset: Average return length as a function of plateau length for the Au/ferrocene/Au junction. Return length is defined as the length required to retract and reform the junction after breaking. All measurements presented in this panel were taken at a bias voltage of 300 mV under cryogenic conditions (4.2 K).

The appearance of a conductance plateau generally indicates the formation of a molecular junction, though its characteristics vary from trace to trace due to unique evolution of each junction. To get statistical behavior, in **Figure 2c**, we have plotted the conductance histogram of an Au atomic junction before (black) and after (blue) insertion of molecule. Sharp peak at 1.0 $G_0$ for both the histograms corresponds to the conductance of single atomic Au constriction[3]. Below 1.0 $G_0$, several wriggles are observed, with two prominent peaks around ~ 0.1 $G_0$ and ~ 0.01 $G_0$ in the blue histogram, which was flat and featureless in absence of the molecule due to tunneling between the electrodes (black histogram). Furthermore, combined analysis of conductance histograms (Supplemental **Figure S3**) and conditional conductance histograms (Supplemental **Figure S4**) at different bias voltages reveal the effect of electric field to the formation of plateaus. While three types are observed at all bias voltages, LG plateaus are more stabilized at higher bias voltages, contrasting with HG plateaus, which are less favored at higher biases.

The evolution of conductance plateaus during stretching can clearly be visualized by the 2D density plots, which are constructed by aligning all the traces to a particular conductance (0.3 $G_0$). **Figure 2d** reveals the characteristic data clouds resembling the tilted 'Z' and 'U'-shaped features for HG and LG plateaus, respectively, as observed for the individual traces. Such a peculiar conductance-displacement plot is so far unknown to us and hint at an unconventional electro-mechanical response of the Au/ferrocene/Au junction. The contrasting evolution of the HG and LG plateaus becomes more apparent in the conditional 2D conductance displacement histograms, as illustrated in Supplemental **Figure S5**. Though three data clouds are clearly visible in **Figure 2d**, multiple faint U-shaped features can be seen beyond 6 Å. The short data cloud denoted by $ in **Figure 2d** is attributed to intermediate plateaus of Type-2. Note that bias dependent formation probabilities of the HG and LG plateaus are also reflected in conductance displacement histogram where U-shaped plateaus intensify at higher bias voltages (Supplemental **Figure S6**).

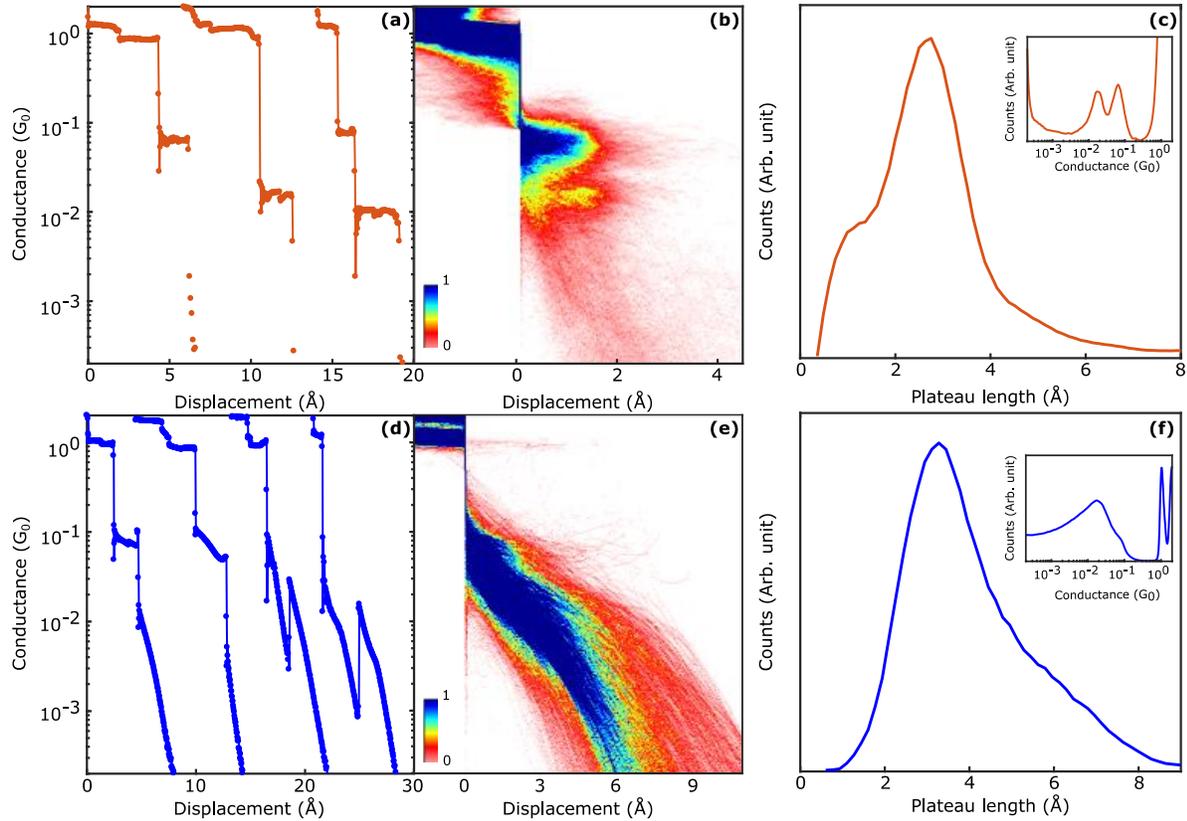

**Figure 3: Effect of temperature and making traces characteristics of the Au/ferrocene/Au junction.** (a) Conductance displacement breaking traces of the same junction, measured at 77 K. Traces are shifted horizontally for better visualization. (b) 2D conductance displacement histogram, prepared from the 9566 conductance displacement traces using 50 bins per decade, recorded at 77K. Conductance of the zero displacement is 0.1 $G_0$. (c) Histogram of plateau lengths, having conductance between 0.3 $G_0$ to 0.001 $G_0$ using 200 bins. Inset: Conductance histogram, using 50 bins per decade based on the same set of traces as used in (b). Measurement presented in figure panel a-c is taken at 100 mV bias voltage. (d) Conductance displacement making traces of the Au/ferrocene/Au junction, recorded on approaching the broken electrodes at 4.2 K. (e-f) 2D conductance displacement histogram (e) and plateau length histogram (f) of the similar making traces. Zero displacement of 2D histogram is taken at 0.5 $G_0$ and conductance segment for plateau length histogram, is 0.3 $G_0$ to 0.0005 $G_0$. Panels d-f are the making counter part of the breaking event as shown in Figure 2 at 300mV bias voltage. Inset: Conductance histogram based on the same set of traces as used in (e) using 50 bins per decade.

We next focus on the length of the molecular plateaus, which are elongated considerably upon stretching (maximum up to ~ 16 Å). The lengths of the molecular plateaus (0.3 $G_0$ < G < 0.0005 $G_0$) are calculated from individual breaking traces and presented as a histogram in **Figure 2e**, showing the distribution of plateau lengths across all traces. A sharp peak at 1.4 Å corresponding to the tiny HG plateaus along with multiple peaks or wriggles up to 15 Å can be seen. Peaks in the length histogram are attributed to different adsorption geometries, as the stretching of the junction can lead to an uneven distribution of atoms and molecules, as well as conformational modifications of the molecules. Interestingly, the average separation between the peaks 1, 2 and 3 of the length histogram is ~ 2.65 Å, closely matching the typical repeated unit observed in Au atomic chains[17,18]. Thus, length of the plateau varies according to the number of atoms involved in the junction, which is reflected as equidistant peaks in the length histogram with peak separation comparable to the interatomic separation of Au atom, a characteristic signature of chain formation[16–18]. Thus, our experimental findings indicate the formation of ferrocene-assisted gold atomic chains. Although atomic chain formation has been reported for various metal-molecule combinations[10–16,19–27], this is, to the best of our knowledge, the first example of an atomic chain guided by an organometallic molecule. Additional verification for molecule-assisted chain formation is obtained by plotting average return length as a function of plateau length (inset of **Figure 2e**). Return length is defined as the distance that two electrodes must retract to reestablish the junction after it is broken. Average return length increases almost linearly with the plateau length which further reestablishes our claim of ferrocene assisted Au atomic chain formation. However, occasionally, traces without chain show short, increasing molecular plateaus in the HG region. Length histograms at other bias voltages display features up to 16 Å, with the proportion of longer plateaus (> 5 Å) decreasing at higher biases (Supplemental **Figure S7**), probably due to Joule heating, vibrational excitations, and current-induced embrittlement[31–35].

To investigate the unusual electro-mechanical response and its link to chain formation, we repeat the experiment at a higher temperature (77 K), where chain formation is less likely due to enhanced thermal energy. Notably, peculiar evolution of molecular plateaus (< 1.0 $G_0$) is absent and the conductance traces (**Figure 3a**) exhibit two flat plateaus at HG and LG regions, which is also evident from the 1D conductance histogram (inset, **Figure 3c**) and 2D conductance-density plots (**Figure 3b**). These two plateaus can form concurrently or independently (Supplemental **Figure S8**). Moreover, plateau length histogram (**Figure 3c**) is dominated by a single peak at ~ 2.8 Å, which eliminates the possibility of molecule assisted chain formation. Thus, absence of chain formation at 77 K results in sharp peaks of the conductance histogram due to reduced structural variation, whereas broad width and fine peaks in the 4.2 K conductance histogram (blue plot, **Figure 2c**) reflect the rich structures of ferrocene-assisted gold chains. Hence, our experiment at 77K demonstrates no chain formation and no unusual electro-mechanical response. For additional data sets, see Supplemental **Figure S9**.

Analyzing the making traces at 4.2 K provides valuable insights, as the atoms involved in the atomic chain collapse to either electrode upon breaking, creating blunt electrode tips during reformation[18]. When pushed, these blunt tips facilitate quantum tunneling and may form either molecular junctions or collapse to metallic contacts. Interestingly, unusual evolution of plateau is missing in push traces. Instead, plateaus are either flat or decaying (**Figure 3d**), as manifested in the density cloud of the conductance displacement histogram (**Figure 3e**). A broad peak in the length histogram (**Figure 3f**) suggests that no chain is involved in the re-formation of molecular junction. However, plateau lengths can exceed 5 Å, as making traces exhibits longer plateaus by accessing more configurations which was masked by the "snap-back effect" in case of breaking process[36,37]. Meanwhile, the conductance histogram of the making traces depicts a sharp peak at ~ 0.02 $G_0$ (Inset of **Figure 3f**). Overall, our experimental observations for Au/ferrocene/Au junction reveal a strong interdependence between unusual evolution of conductance plateaus (either increasing or U-shaped features) and molecule assisted chain formation. It is worth highlighting that the earlier experiment involving ferrocene and silver electrodes, where atomic chains are absent, showed no evidence of such unconventional electro-mechanical behavior[38,39].

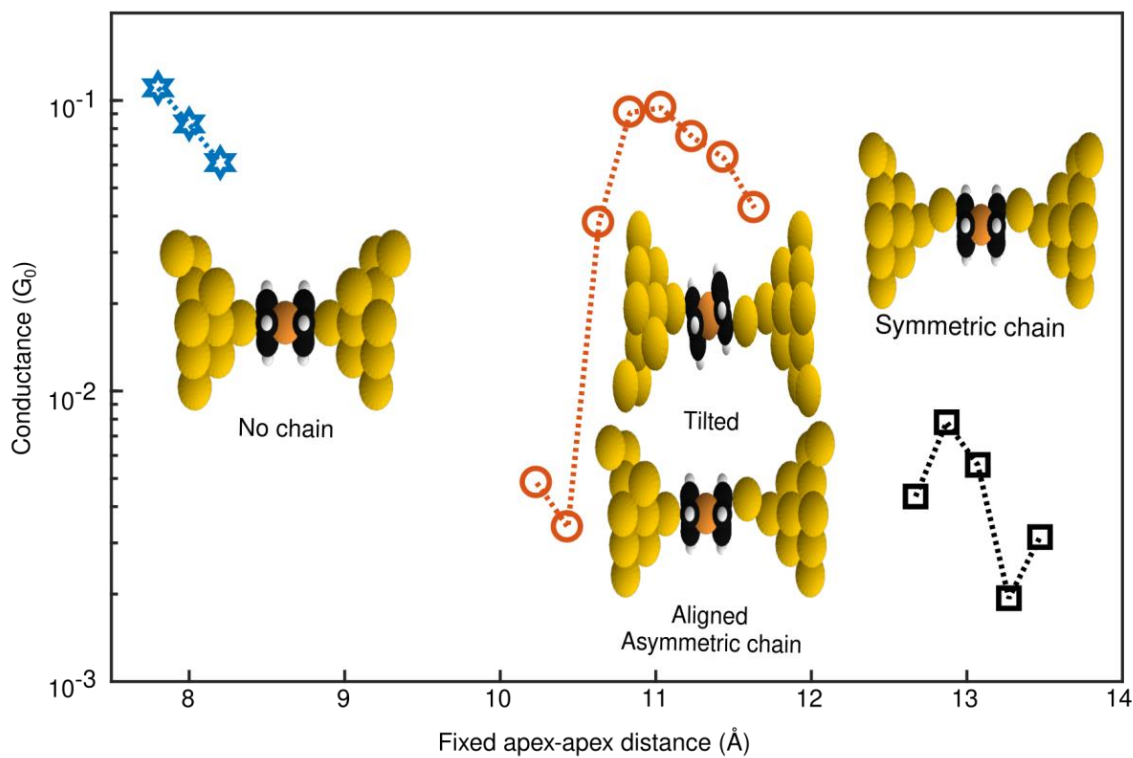

**Figure 4: Ab-initio results.** Conductance as a function of the (pyramid) apex-to-apex distance of Au/ferrocene/Au junction. For junctions without the Au chain, the ferrocene molecule does not tilt within the investigated geometries and the conductance decreases monotonously. For junctions bound to the chain, reorientation and non-monotonous conductance behavior is observed.

To understand the influence of molecule assisted chain formation on the electromechanical response of Au/ferrocene/Au junction, we have performed DFT-based first-principles calculations for a set of Au/ferrocene/Au junctions. Ferrocene, which can be oriented either parallel or perpendicular to atomistic electrodes[38–41], is found to support stable chains only in the parallel orientation (see **Section 1** of the Supplementary Information for details). The conductance of parallel configurations is analyzed for three cases: no chain atoms in the electrodes (No Chain), chain atoms on one side (Asymmetric Chain), and chain atoms on both sides (Symmetric Chain), as shown in **Figure 4**. In line with the experimental findings, unusual electro-mechanical responses are observed for ferrocene connected via Au chains, particularly in the asymmetric chain configuration, whereas conventional decaying behavior is noted in the absence of chains. The underlying mechanism is then explored by calculating transmission functions for the asymmetric chain geometry at varying electrode separations (**Figure 5a**). Interestingly, beyond 10.435 Å, a significant broadening of the LUMO peak occurs, resulting in increased conductance. This change in broadening correlates with the tilting of the ferrocene molecule, modifying the metal-molecule contact site from the center of the cyclopentadienyl (Cp) ring to a single carbon atom. In **Figure 5b,c,** this recoupling is seen at the left apex.

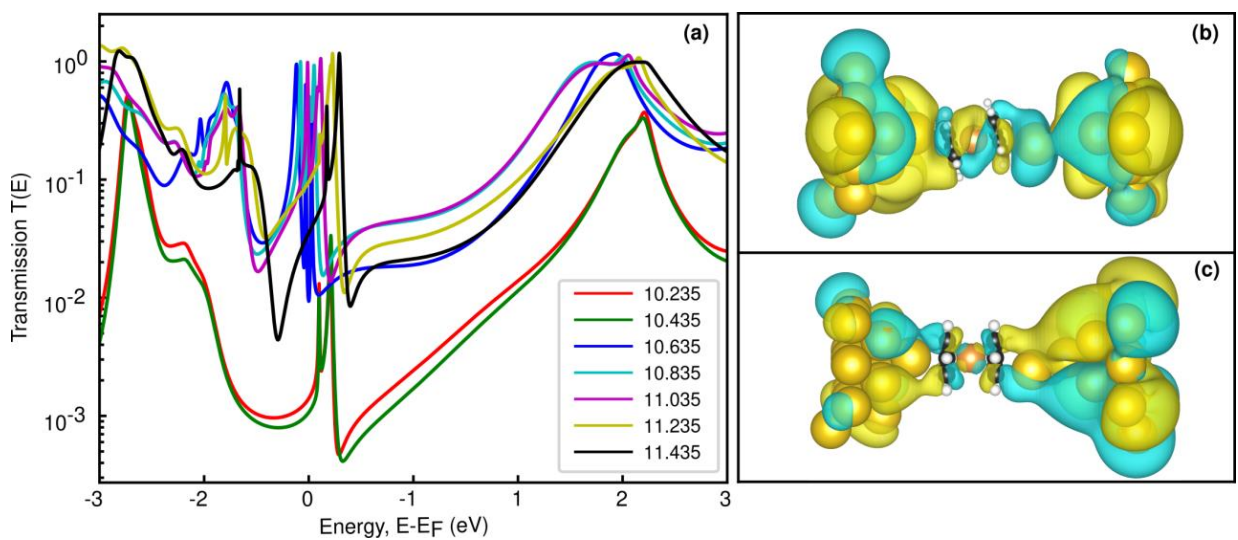

**Figure 5: Transmission functions and transmitting orbitals near the Fermi level.** (a) Transmission functions of the asymmetric chain geometries for different (pyramid) apex-to-apex distances. Inset indicates the pyramid apex to apex distance in Å corresponding to each color. At around 10.5 Å, a reorientarion of the ferrocene occurs, represented by the structures in b (for 11.035 Å) and c (for 10.635 Å). The reorientation causes a dramatic change in the conductance. **(b-c)** show the transmitting orbital with energy close to the Fermi level and nodal structure of LUMO of the ferrocene in isolation.

The reorientation abruptly changes the hybridization of LUMO with the conduction band of Au, which can be understood as follows. The LUMO on the Cp rings consist predominantly of the $p_z$ orbitals whose sign changes twice upon a complete rotation around the ferrocene axis (see Supplementary Information of reference 40). Therefore, when the apex binds to all 5 carbons, the LUMO can only hybridize with a $p_x$ orbital of the apex, where x is perpendicular to the transport axis (z). On the other hand, when the apex binds to the single C, the local symmetry allows the LUMO to hybridize both with s and $p_z$ orbitals of the apex. It is this conformation that allows the LUMO to hybridize more efficiently with the conduction band, consequently increasing the broadening. A similar mechanism was previously described[42], where the orbital symmetry differentiated the resonance broadenings of two molecules. Here, this mechanism occurs in a single molecule upon conformation change. We note that below 10.5 Å the unoccupied transmission resonances do not reach unity because of a pronounced left-right asymmetry, further negatively impacting the conductance. All transmission functions and junction geometries are available online[43].

**Conclusions:**

We uncover unconventional electro-mechanical behavior of ferrocene molecular junctions, enabled by the formation of gold atomic chains at cryogenic temperatures (4.2 K). Experimental results reveal atypical conductance plateaus, characterized by rising or 'U'-shaped features during the breaking events at 4.2 K. In contrast, such plateaus and chain formation are absent during the corresponding making events or at higher temperature (77 K). These anomalies in conductance plateaus are attributed to the ferrocene-assisted formation of gold chains. DFT-based calculations corroborate the findings, showing that the microscopic conformation of ferrocene and the presence of gold chains significantly affect the conductance by modifying metal-molecule contact points and enhancing orbital overlap. Overall, our study underscores the crucial role of molecule-assisted chain formation in driving unconventional electromechanical responses and proposes organometallic molecules as promising platforms for developing mechano-sensitive molecular devices.

**ACKNOWLEDGMENTS**

B.P. acknowledges the support from DST-Inspire fellowship, Government of India (Inspire code IF170934), and A.N.P. acknowledges the funding from the Science and Engineering Research Board, Department of Science and Technology, Government of India (Grant No. CRG/2020/004208). B.P and A.N.P acknowledges Debayan Mondal and Priya Mahadevan for fruitful discussions. Š.M. acknowledges the computational resources provided by e-INFRA CZ project (ID:90254), supported by the Ministry of Education, Youth and Sports of the Czech Republic. R.K. acknowledges the support of Czech Science Foundation (GAČR) through grant 22-22419S.

# Unconventional electro-mechanical response in Ferrocene assisted gold atomic chain

Biswajit Pabi[a,$,*], Štěpán Marek[b,c], Tal Klein[d], Arunabha Thakur[e], Richard Korytár[b] and Atindra Nath Pal[a]*

[a]Department of Condensed Matter and Materials Physics, S. N. Bose National Centre for Basic Sciences, Sector III, Block JD, Salt Lake, Kolkata 700106, India.
[b]Department of Condensed Matter Physics, Faculty of Mathematics and Physics, Charles University, CZ-121 16 Praha 2, Czech Republic.
[c]Institute of Theoretical Physics, University of Regensburg, Universitätsstraße 31, D-93053, Regensburg, Germany.
[d]Department of Chemical and Biological Physics, Weizmann Institute of Science, Rehovot 7610001, Israel.
[e]Department of Chemistry, Jadavpur University, Kolkata-700032, India

[$]Present address: Institut für Experimentelle und Angewandte Physik, Christian-Albrechts-Universität, 24098 Kiel, Germany.

*Email: biswajitpabi1@gmail.com, atin@bose.res.in

**Contents:**

1. **Methodology:**

    (i) **Experimental methodology**

A mechanically controllable break junction is used to form the molecular junction where a notched Au wire (99.997%, 0.1 mm, Alfa Aesar) is fixed on a flexible substrate (1 mm phosphor bronze, 100 µm Kapton foil). The setup, placed in a vacuum chamber and cooled down to ~ 4.2K, uses a three-point bending mechanism to gradually stretch the wire to atomic scale. Fine bending is controlled by a piezoelectric actuator (PI P-882 PICMA), driven by a 24-bit DAQ card (PCI 4461, National Instruments) and a piezo driver (SVR 150/1, Piezomechanik). Bending the substrate breaks the Au wire into two segments with freshly exposed atomic tips, which act as electrodes in an ultraclean cryogenic environment. Molecular junctions are formed by repeatedly breaking and reforming the atomic contact between the tips while sublimating ferrocene molecules from a heated source. Recrystallized ferrocene (purchased from LOBA Chemie) from dichloromethane (DCM) was used directly for conductance measurements.

Conductance-displacement traces are obtained by measuring the dc conductance during the breaking or making processes. A constant bias voltage from the DAQ card, reduced by a factor of 10 to improve signal-to-noise ratio, is applied across the junction. The resulting current output from the junction is amplified by an I/V preamplifier (Femto DLPCA-200), and the amplified signal is recorded by the same DAQ card. The dc conductance is calculated by dividing the measured current by the applied voltage. Inter-electrode displacement is estimated based on the dependence of tunneling current on electrode separation, following a standard procedure[1].

   (ii) **Theoretical methodology**

Our computational results start by determining a set of geometrically optimized atomic coordinates. We choose a ferrocene in the parallel and perpendicular geometry (**Figure 4**) with pyramidal gold electrodes of 3 layers and two ad-atoms. The distance between the apex atoms of the pyramid is varied and for very large distances, ad-atoms from the electrodes are moved beyond the apex of the electrode to serve as the gold atomic-chain atoms. These initial geometries are relaxed by geometrical optimization based on Kohn-Sham DFT[2] in the PBE functional[3] approximation with disp4 correction[4], as implemented in the TURBOMOLE[5] code. For geometry optimization, the threshold on energy change was $10^{-5}$ H, on gradient the threshold was $10^{-2}$ a.u. For the self-consistent field cycle, the threshold on total energy change was $10^{-7}$ H. The Gaussian basis set used was def2-SVP[6] in the first step of the optimization and def2-TZVP in the second step, both used with pseudopotentials[7] for gold atoms.

For the perpendicular starting geometry, geometry optimization led to the breaking of the junction for nearly all distances of the apex atoms investigated when chain atom was present. Specifically, we investigated apex-apex distances from 11.47 Å to 13.27 Å for symmetric chain (10 geometries in total) and distances from 9.435 Å to 11.035 Å for the asymmetric chain (4 geometries in total). In all but two cases (13.07 Å for symmetric chain and 11.035 Å for the asymmetric chain), the ferrocene molecule drifted from the central position to position on a single electrode, and no unexpected behavior in the transmission function/conductance was observed. We therefore focus on the parallel geometry in the main text and in the rest of this text.

The transmission function was determined from the Kohn-Sham orbitals and eigenvalues (from the def2-TZVP basis) using the non-equilibrium Green's function method[8], assuming infinitesimal bias and balancing the absorbing boundary conditions by requiring charge neutrality to within $10^{-7}$ e. The absorbing boundary conditions were applied only to the outermost layer of the gold electrodes and corresponded to bandwidth of 0.1 H. The resulting transmission function was integrated with the difference of Fermi distribution functions, with temperature set to 4.2 K. Transport wavefunctions were visualized using VESTA[9].

2. **Characterization of ferrocene**

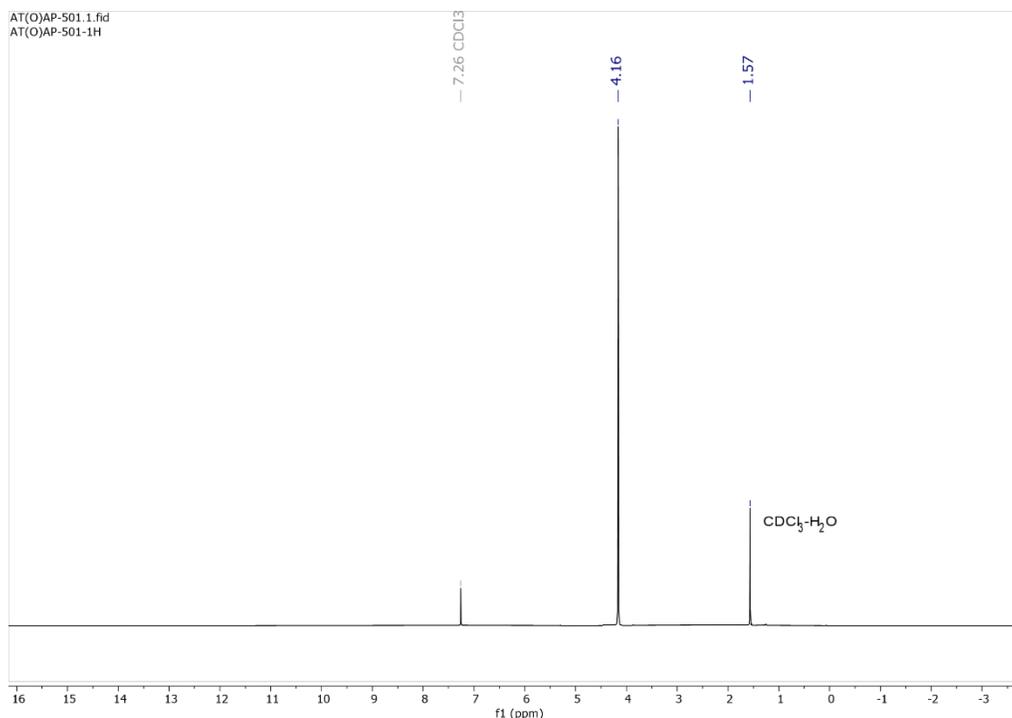

**Supplemental Figure S1: NMR spectra of studied ferrocene.** $^1$H NMR spectrum of ferrocene in CDCl$_3$.

## 5. Additional analysis of the low temperature (4.2 K) data.

### (i) Conductance histogram without data selection.

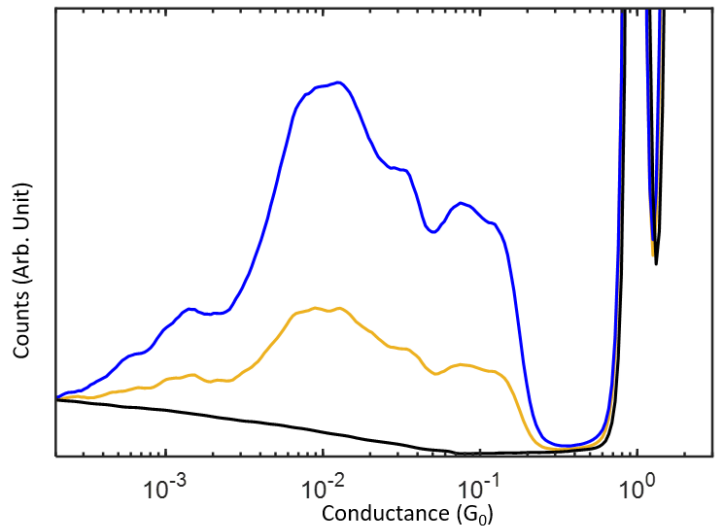

**Supplemental Figure S2: Conductance histogram at 300 mV bias voltage.** Conductance histogram of Au/ferrocene/Au junction considering total traces i.e. without data selection (yellow), traces with molecular features i.e. molecular traces (blue) and traces without molecular features i.e., tunneling (black). The data presented here is taken at 300 mV bias voltage.

### (ii) Conductance histogram at different bias voltages.

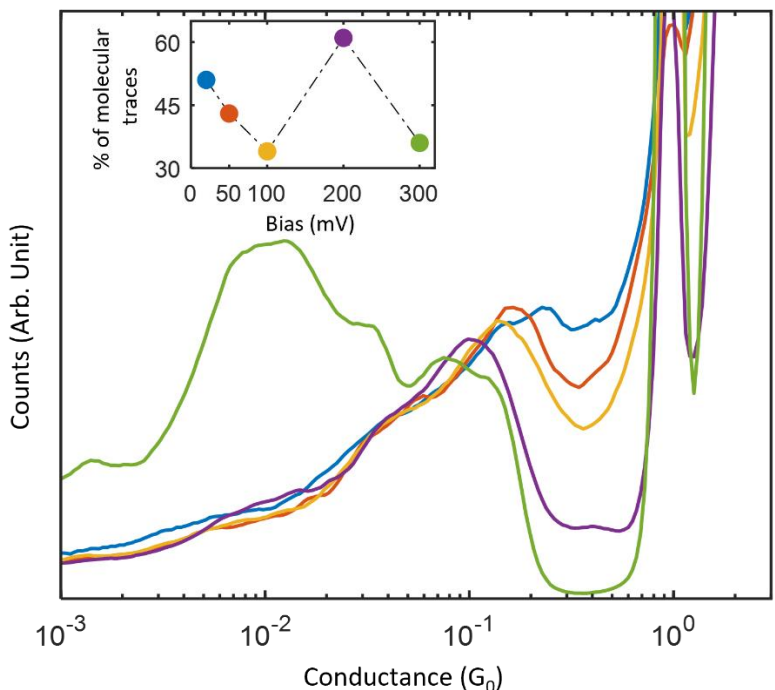

**Supplemental Figure S3: Conductance histogram for different bias voltages.** Conductance histogram of Au/ferrocene/Au junction for bias voltages of 20 mV, 50 mv, 100 mv, 200 mV and 300 mV. For color corresponding to the bias voltage, please see the inset which plots the percentage of molecular traces as a function of bias voltage.

**Supplemental table S1:** Number and percentage of molecular traces at different bias voltage.

| Bias Voltage | Total traces | Molecular traces | Percentage of molecular traces (%) |
|---|---|---|---|
| 300 | 19948 | 7099 | 36 |
| 200 | 20000 | 12131 | 61 |
| 100 | 19995 | 6692 | 34 |
| 50 | 20000 | 8499 | 43 |
| 20 | 16812 | 8482 | 51 |

(iii) Conditional analysis of three types of traces at various bias voltages.

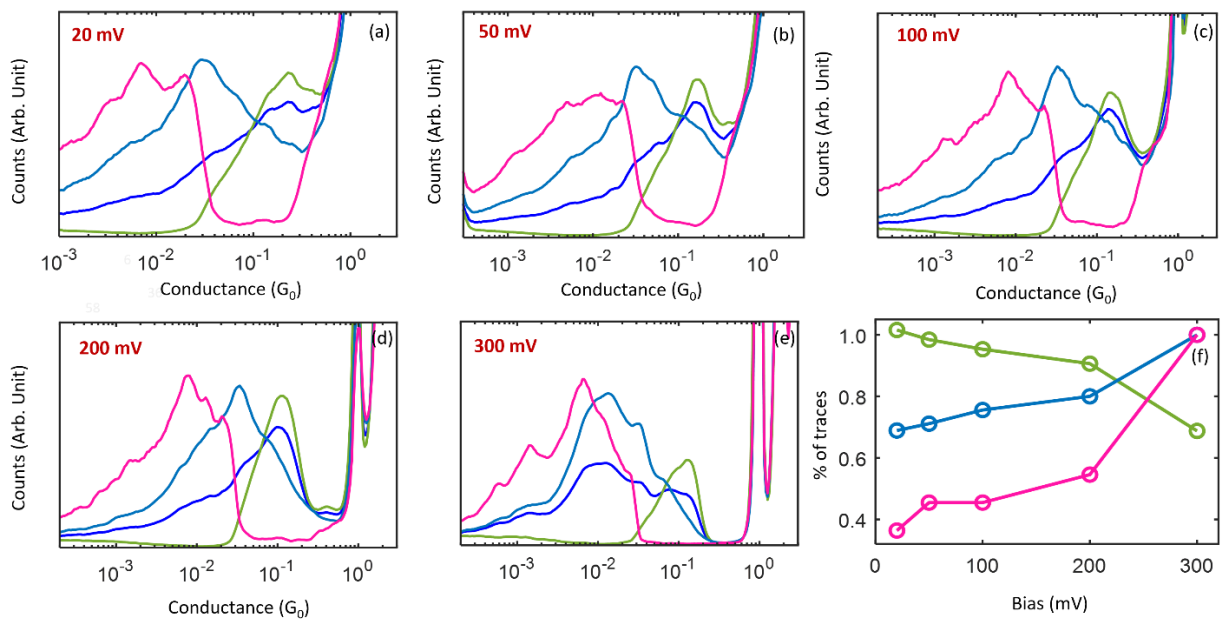

**Supplemental Figure S4: Conductance histogram of Type-1, 2 and 3 traces for different bias voltages. (a-e)** Conductance histogram of three types of traces, mentioned in **Figure 2** where green blue and magenta correspond to Type-1, 2 and 3, respectively. Respective bias voltage is mentioned on the top-left. **(f)** Plotting of normalized percentage of Type-1, 2 and 3 traces for individual bias voltages.

**Supplemental table S2:** Number and percentage of Type-1, Type-2 and Type-3 traces at the considered bias voltage.

| Bias Voltage | Type-1 | | Type-2 | | Type-3 | |
|---|---|---|---|---|---|---|
| | Number of traces | Percentage (%) | Number of traces | Percentage (%) | Number of traces | Percentage (%) |
| 300 | 3113 | 44 | 3198 | 45 | 788 | 11 |
| 200 | 7067 | 58 | 4327 | 36 | 737 | 6 |
| 100 | 4070 | 61 | 2312 | 34 | 310 | 5 |
| 50 | 5314 | 63 | 2737 | 32 | 448 | 5 |
| 20 | 5427 | 64 | 2649 | 30 | 406 | 4 |

**(iv)**      **2D conductance displacement histogram of three types of traces.**

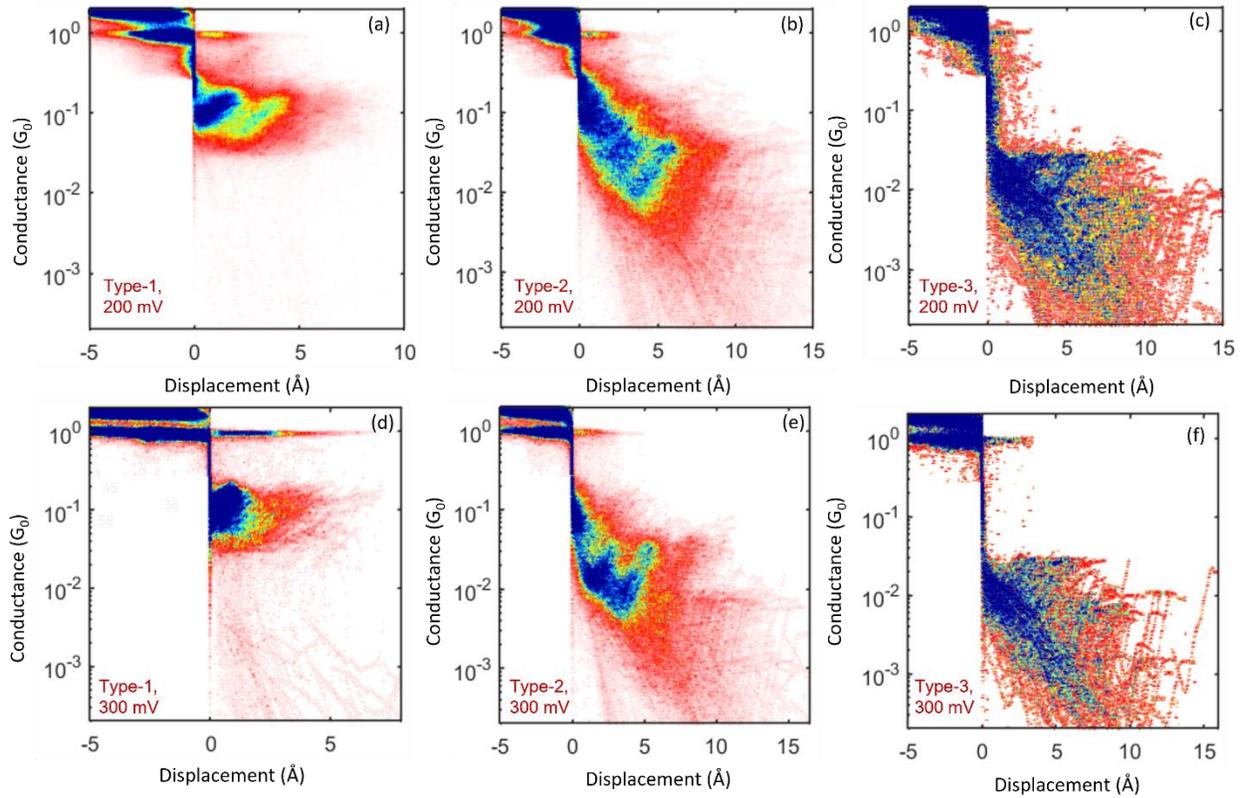

**Supplemental Figure S5: 2D Conductance displacement histogram for three types of traces (1, 2, 3). (a-f)** 2D Conductance-displacement density plot of three types of traces of Au/ferrocene/Au junction for 200 mV (a-c) and 300 mV (d-f) bias voltage. Here, the conductance of the zero displacement is 0.3 $G_0$. Corresponding bias voltage along with the types are mentioned in the bottom-left corner.

**(v)  2D conductance-displacement density plot at other bias voltages.**

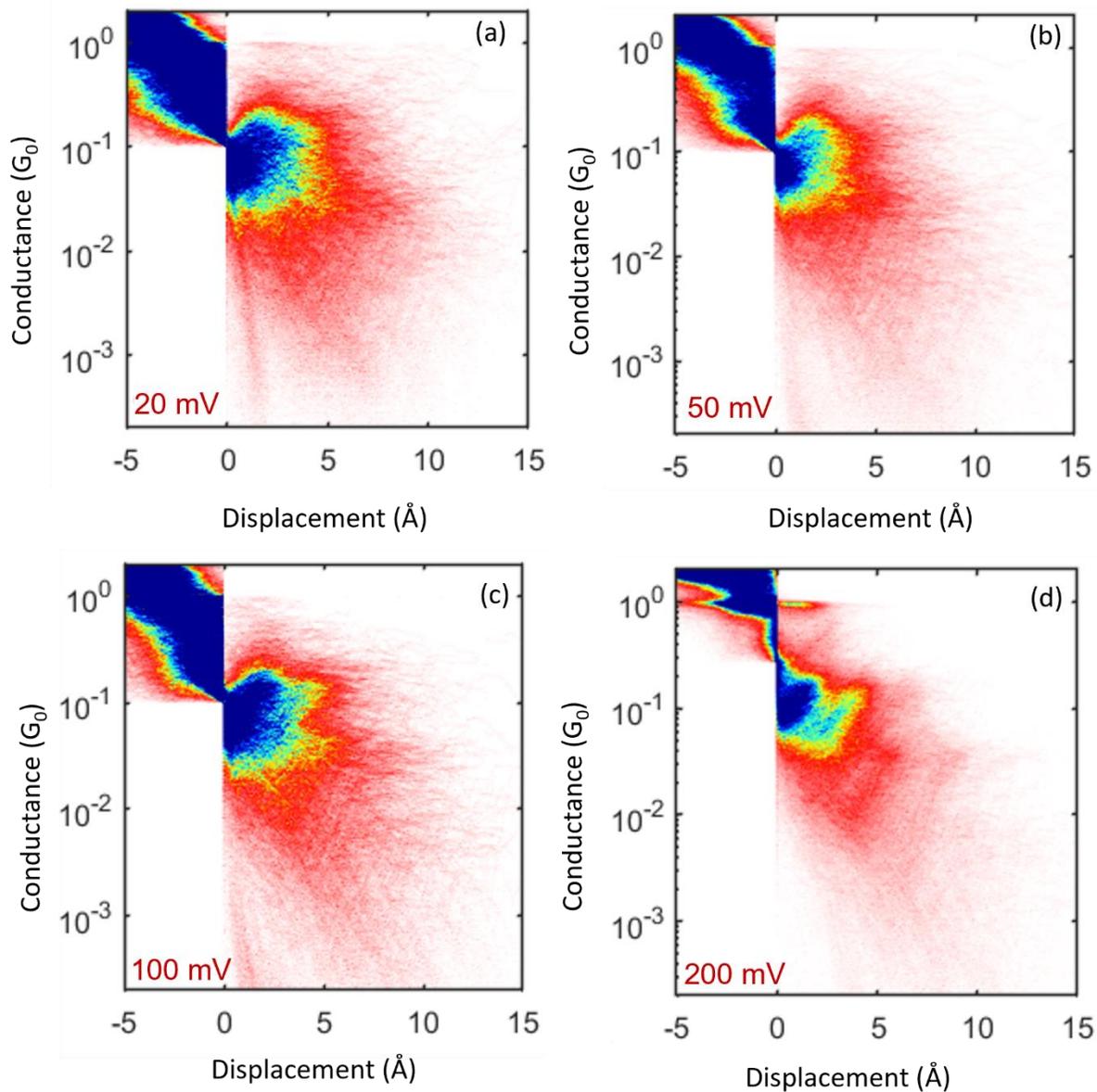

**Supplemental Figure S6: 2D Conductance displacement density plot for different bias voltages. (a-d)** 2D Conductance displacement histogram of Au/ferrocene/Au junction for bias voltage 20 mV, 50 mv, 100 mv, 200 mV and corresponding bias voltage is mentioned in the bottom-left corner. 0.3 $G_0$ is considered as the conductance of zero displacement.

**(vi)    Plateau length histogram at different bias voltages.**

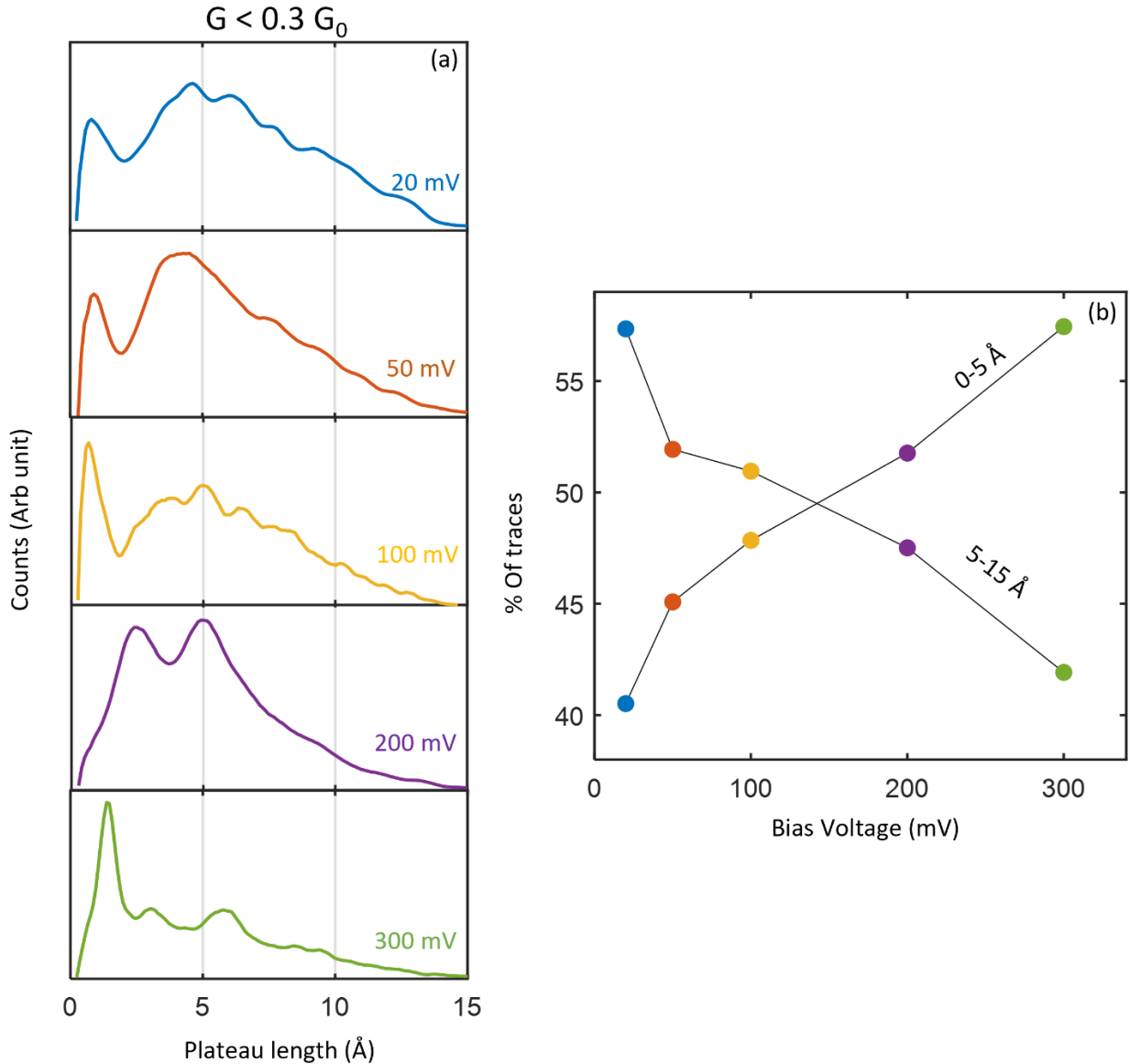

**Supplemental Figure S7: Plateau length histogram for different bias voltage. (a)** Histogram of plateau lengths from zero displacement (0.3 $G_0$) to breaking (0.0005 $G_0$) of Au/ferrocene/Au junction for bias voltage 20 mV, 50 mv, 100 mv, 200 mV, 300 mV and corresponding bias voltage is mentioned in the bottom-right corner. **(b)** Percentage of traces with plateau length 0-5 Å and 5-15 Å as a function of bias voltage, following the same colors.

## 3. Additional data and analysis from the measurements at 77 K.

### (i) Three types of conductance traces and their histogram.

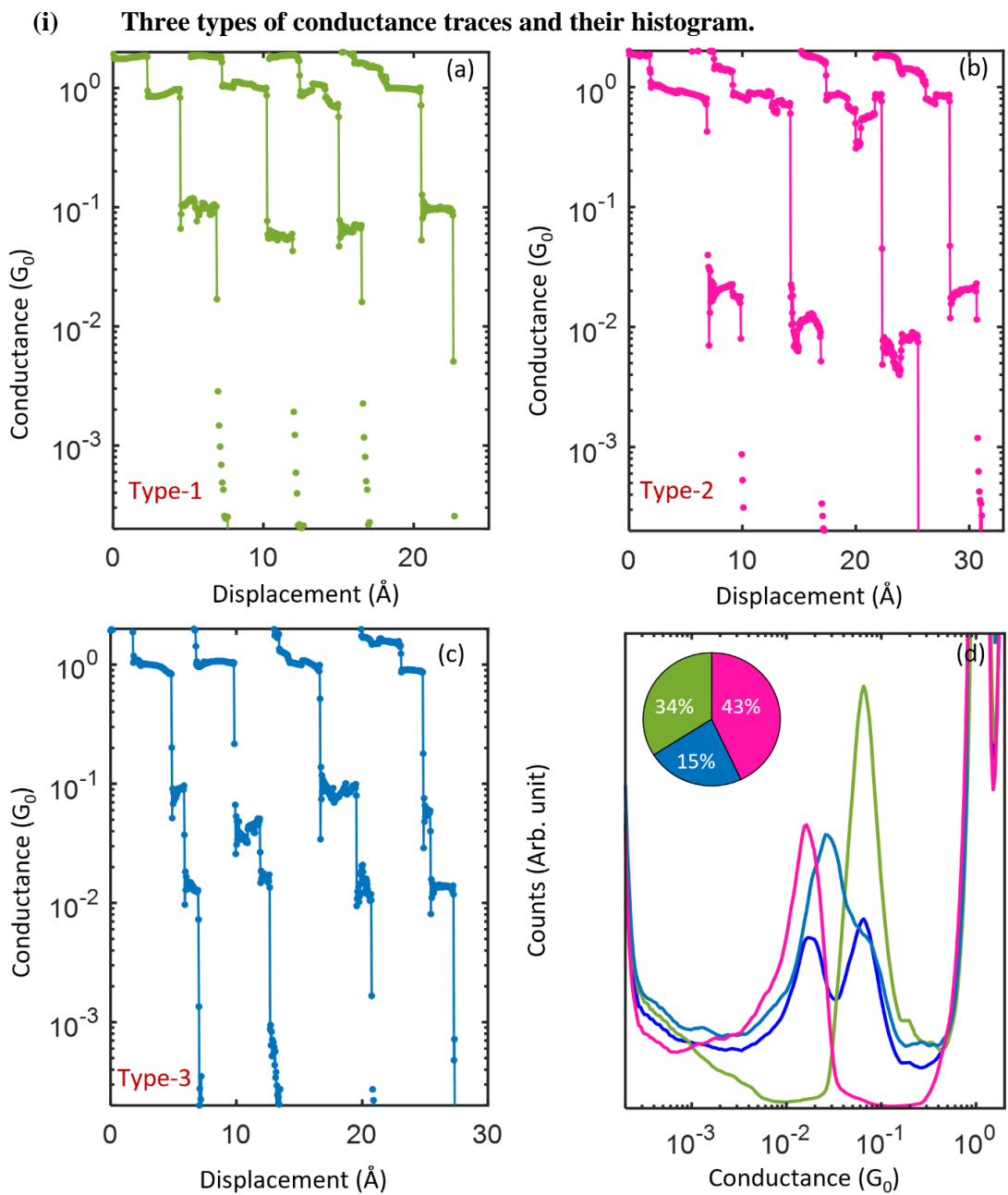

**Supplemental Figure S8: Conductance characterization of Type-1, 2 and 3 traces recorded at 77 K. (a-d)** Typical conductance displacement breaking traces (a-c) and conductance histogram (d) of three types of traces, captured at 77 K for Au/ferrocene/Au junction. Green, blue and magenta corresponds to Type-1, 2 and 3 and mentioned at the bottom left corner. Three types are defined based on the appearance of plateaus either at ~ 0.1 $G_0$ or at 0.01 $G_0$ or at both. The pie chart of inset (d) represents the percentage of Type-1, 2 and 3 traces out of total molecular traces. The measurement of that figure panel is performed at a bias voltage of 100 mV.

**(ii)    Data from another set of measurements at 77 K.**

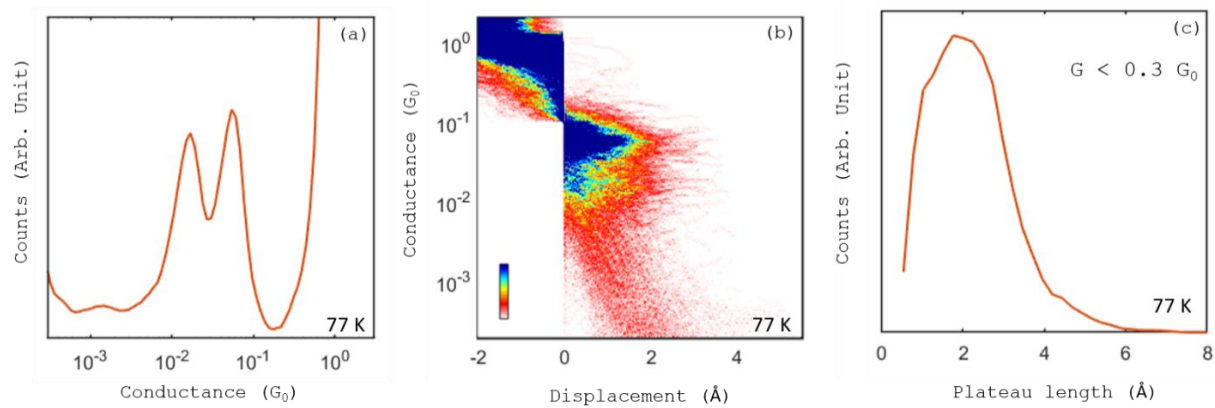

**Supplemental Figure S9: Conductance characterization of another set of measurements at 77 K.** (a-c) Conductance histogram (a), 2D conductance displacement histogram (b) and plateau length histogram (c), prepared from the breaking traces of Au/ferrocene/Au junction at 77 K. Plateau length histogram is based on the plateaus with conductance value between 0.3 $G_0$ and 0.001 $G_0$. Measurements, shown in that figure panel, are carried out with 100 mV bias voltage.